\documentclass{aa}
\usepackage{psfig}
\begin{document}

\title{Detection of a warm molecular wind in DG Tauri\footnote{Based on data collected at the Subaru Telescope, which is operated by National Astronomical Observatory of Japan}}


\author{M. Takami\inst{1}     \and
        A. Chrysostomou\inst{1} \and
        T.P. Ray \inst{2}       \and
        C. Davis \inst{3}       \and
        W.R.F. Dent  \inst{4}       \and
        J. Bailey\inst{5}       \and
        M. Tamura\inst{6}       \and
        H. Terada\inst{7}}

\institute{Department of Physical Sciences, University of
           Hertfordshire, College Lane, Hatfield, Herts AL10 9AB, UK
           \and
           School of Cosmic Physics, Dublin Institute for Advanced Studies, 5 Merrion Square, Dublin 2, Ireland
           \and
           Joint Astronomy Centre, 660 North A'ohoku Place, University Park, Hilo, Hawaii 96720, USA
           \and
           UK Astronomy Technology Centre, Royal Observatory, Blackford Hill, Edinburgh EH9 3HJ, UK
           \and
           Anglo-Australian Observatory, PO Box 296, Epping, NSW 1710, Australia
           \and
           National Astronomical Observatory of Japan, Osawa, Mitaka, Tokyo 181-8588, Japan
           \and
           Subaru Telescope, 650 North A'ohoku Place, Hilo, Hawaii 96720, USA
}

\offprints{Michihiro Takami, \email{takami@star.herts.ac.uk}}

\date{Received 17 October 2003 / Accepted 28 November 2003}

\abstract{
We detect near-infrared H$_2$ emission in DG Tau using the Infrared Camera and Spectrograph (IRCS) on the 8.2-m SUBARU telescope.
The spectra obtained along the jet axis show that the centroidal position of the 1-0 S(1) emission is offset by 0.2'' from the star towards the jet, while those obtained perpendicular to the jet axis show a marginal extension, indicating that the emission line region
has a typical width of $\sim$0.6''. Their line profiles show a peak velocity of $\sim$15 km s$^{-1}$ blueshifted from the systemic velocity.
These results indicate that the emission originates from a warm molecular wind with a flow length and width of $\sim$40 and $\sim$80 AU, respectively. The line flux ratios ($I_{1-0 S(0)}/I_{1-0 S(1)}$ and an upper limit for $I_{2-1 S(1)}/I_{1-0 S(1)}$) suggest that the flow is thermalized at a temperature of $\sim$2000 K, and is likely heated by shocks or ambipolar diffusion. The observed velocity and spatial extension suggest that the H$_2$ and forbidden line emission originate from different components of the same flow, i.e., a fast and partially ionised component near the axis and a slow molecular component surrounding it. Such a flow geometry agrees with model predictions of magneto-centrifugal driven winds.
}

\titlerunning{a warm molecular wind in DG Tauri}

\authorrunning{Takami et al.}

\maketitle

\section{Introduction}
Understanding the mechanism of mass accretion and driving of jets/winds is one of the most important key issues for star formation theories. There is growing evidence that such outflows are powered by disk accretion. Disk accretion rates estimated through UV excess measurements are correlated with outflow signatures such as forbidden line luminosities (Hartigan, Edwards, \& Ghandour 1995). Recent high-resolution observations have suggested that the jet appears to rotate around the flow axis (Davis et al. 2000; Bacciotti et al. 2002), supporting the scenario that the jet removes the excess angular momentum from the circumstellar disk, allowing the accretion to occur.

Among the models proposed to date, magneto-centrifugal wind models seem the most promising as the mechanism for driving the jet/wind. 
According to these models, magnetic and centrifugal forces act together to launch the jet/wind along magnetic field lines, from a region of the disk between $\le$0.1 (``X-wind'' --- see Shu et al. 2000) and a few tens of AU (``disk wind'' --- see, e.g., K\"onigl \& Pudritz 2000; Ferreira 1997). The magnetic field lines act as solid wires up to the ``Alfv\'en surface'', located at less than 20 AU from the disk. These accelerate the flow particles outwards and upwards in a 'bead-on-a-wire' fashion, simultaneously removing the angular momentum from the accretion disk. Although present high-resolution facilities cannot resolve such kinematics in the innermost region, the models seem to explain some key observed properties, including (1) the observed mass ejection per mass accretion rates ($\sim$0.1, see e.g., Calvet 1997; Richer et al. 2000), (2) motion in the jet consistent with rotation around the axis (Davis et al. 2000; Bacciotti et al. 2002), and in the same sense as the circumstellar disk or molecular envelope (Wiseman et al. 2001; Testi et al. 2002).

These models also predict the presence of slow and poorly collimated streamlines surrounding a well-collimated jet. Although atomic lines at UV-to-IR wavelengths have been powerful tools to study the jet/wind over decades (see e.g., Eisl\"offel et al. 2000a and Hartigan et al. 2000 for reviews), they may originate from only a limited fraction of the entire flow close to the jet axis (e.g., Li \& Shu 1996; Cabrit, Ferreira, \& Raga 1999; Shang et al. 2002). The presence of an ``unseen'' wide-angled component has been suggested by the presence of a cavity in the molecular core/envelope (see Davis et al. 2002, and references therein), and a broad range of morphology of molecular bipolar outflows (see e.g., Lee et al. 2000, 2001). In addition, Davis et al. (2002) and Saucedo et al. (2003) revealed the presence of shock-heated H$_2$ along the cavity wall in some young stellar objects (YSOs), presumably due to interaction with a wide-angled wind. Direct detection of such a flow component is desired to test the validity of the magneto-centrifugal wind scenario, to investigate the angular transfer problem in detail, and to understand the nature of molecular outflows.  

We report the detection of near-infrared H$_2$ emission in DG Tau, which originates from a warm molecular wind close to the star. The object is one of the most active T Tauri stars known, and exhibits an energetic jet like Class I protostars (e.g., Eisl\"offel \& Mundt 1998; Davis et al. 2003). Since the object is not heavily embedded like Class I protostars ($A_V$=2.2 --- Muzerolle, Calvet, \& Hartmann 1998), it has been observed as one of the best targets for studying outflowing activities close to the star (see, e.g., Solf \& B\"ohm 1993; Kepner et al. 1993; Bacciotti et al. 2000, 2002; Pyo et al. 2003). In particular, Bacciotti et al. (2000, 2002) observed the forbidden line outflow using the {\it Hubble Space Telescope}, and revealed the presence of an "onion-like" kinematic structure, i.e., a continuous bracketing of the higher speed central flow within the lower speed, less collimated, broader flow.

In this paper, we show that the warm molecular wind appears as a part of the ``extensive'' onion-like kinematic structure observed in the forbidden line outflow. The rest of the paper is organised as follows: we describe the details of the observations and results in Sect. 2; we compare our results with previous observations of near-infrared H$_2$ emission in other YSOs, and investigate the driving, location of the launching region, and excitation/heating of the warm molecular wind in Sect. 3; we then give our conclusions in Sect. 4.
Throughout the paper, we adopt a distance to the object of 140 pc based on {\it HIPPARCOS} observations (Wichmann et al. 1998).

\section{Observations and Results}
Observations were made on the night of 2002 November 25 at the SUBARU 8.2-m
telescope using the Infrared Camera and Spectrograph IRCS (Tokunaga et al. 1998; Kobayashi et al. 2000).
The echelle grating mode with a 0.3'' wide slit provides a spectral resolution of $\sim$10$^4$.
The pixel scale of 0.06'' provides good sampling of the seeing profile (0.7--0.8'' during the observations), thereby allowing us to investigate the spatial structure of the emission line region at sub-arcsecond scales. The spectra were obtained along, and perpendicular to, the jet axis (P.A.=226$^\circ$ --- Bacciotti et al. 2000) with a total integration of 480 s for each direction.
The cross disperser was set to observe the following wavelength coverages simultaneously: 1.958--2.011 (order=28), 2.031--2.085 (27), 2.109--2.165 (26), 2.193--2.251 (25), 2.285--2.344 (24), and 2.384--2.446 $\mu$m (23).
In addition to the target, A-type bright standards were observed at similar airmasses to correct for telluric absorption. The flat fields were made by combining many exposures of the spectrograph illuminated by a halogen lamp.

The data were reduced using the KAPPA and FIGARO packages provided by Starlink. We obtained position-velocity diagrams via standard reduction processes: i.e., dark-subtraction, flat-fielding, removal of bad pixels, correcting for curvature in the echelle spectra, wavelength
calibration, correcting for telluric absorption, and night-sky
subtraction. Wavelength calibrations were made by identifying OH airglow and telluric absorption features. For the templates, we used the UKIRT OH line list\footnote{http://www.jach.hawaii.edu/JACpublic/UKIRT/astronomy \\/calib/oh.html} and model spectra provided by ATRAN (Lord 1992). The accuracy of wavelength calibration is better than half a pixel on the detector, corresponding to 4 km s$^{-1}$. The systemic motion of the object was calibrated based on previous observations of Li I $\lambda$6707 photospheric absorption ($V_{\rm Hel}$=16.5 km s$^{-1}$, Bacciotti et al. 2000).

We detected H$_2$ 1-0 S(1) (2.1218 $\mu$m), 1-0 S(0) (2.2233 $\mu$m), 1-0 Q(1) (2.4066 $\mu$m) and 1-0 Q(3) (2.4237 $\mu$m) emission in our cross-dispersed spectra. Since the latter two lines severely suffer from telluric absorption, we use the former two lines and the upper limit of the 2-1 S(1) (2.2477 $\mu$m) flux for the rest of the paper. Fig. 1 shows the position-velocity (P-V) diagram of the 1-0 S(1) emission, and its spatial distribution integrated over $-40$ to 10 km s$^{-1}$. Although we did not fully resolve the emission line region, 1-0 S(1) emission still exhibits a distribution different from the continuum emission. In the spectra obtained along the jet, the position of the 1-0 S(1) emission is displaced from the continuum position by $\sim$0.2'' towards the jet. Those perpendicular to the jet show broader and slightly asymmetric distribution about the continuum. 
To investigate the spatial scale of the emission line region here, we convolved the seeing profile at the continuum with a Gaussian, and plot it in Fig. 2 together with the distribution of the 1-0 S(1) emission. The figure shows that 1-0 S(1) emission has a typical angular scale of $\sim$0.6'' perpendicular to the jet.
The 1-0 S(0) line is much fainter than 1-0 S(1), and its spatial distribution is the same as that of 1-0 S(1) within the uncertainty of the measurement. 

\begin{figure*}
\hspace*{1cm}
    \psfig{file=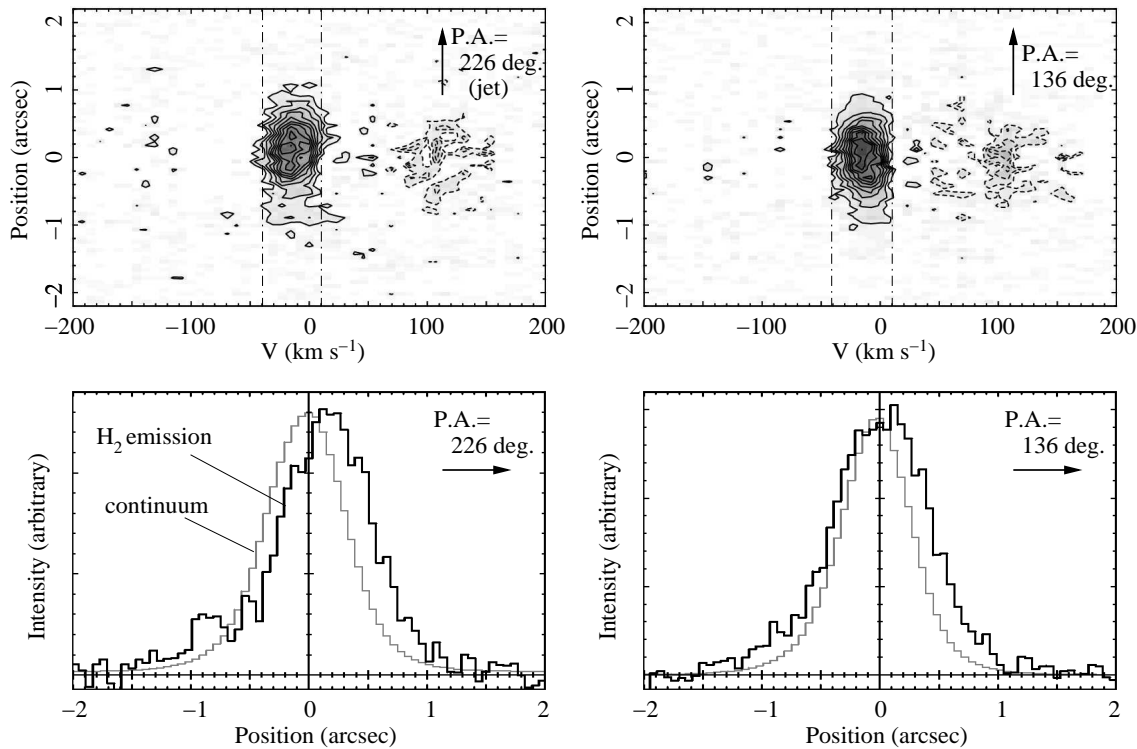,clip=,width=15cm}
  \caption{(top) Continuum-subtracted position-velocity diagrams of 1-0 S(1) emission along and perpendicular to the jet axis. The contour spacings are 10\% of the peak intensity.
The subtracted continuum flux is measured closest to the dot-dashed lines. The continuum flux is not constant along the wavelength due to a number of shallow absorption features, and this causes residual after continuum subtraction at $\sim$100 km s$^{-1}$.
(bottom) Intensity distribution of 1-0 S(1) integrated between dot-dashed lines in the P-V diagrams.
The grey histogram shows the seeing profile observed at the continuum. }
  \label{10S1}
\end{figure*}

\begin{figure}
  \begin{center}
    \hspace*{0.2cm}
    \psfig{file=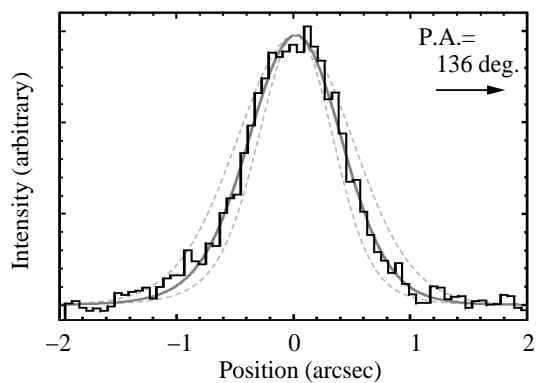,clip=,width=7cm,height=5cm}
  \end{center}
  \caption{Same as the bottom right of Fig. 1, but the seeing profile measured at the continuum is convolved with a Gaussian having a FWHM of 0.6'' (grey solid), 0.3'' and 0.9'' (grey dashed), respectively. The centroidal position of the convolved profiles are adjusted to fit the 1-0 S(1) intensity emission (black solid).}
  \label{profiles}
\end{figure}

Fig. 3 shows the spectra at 1-0 S(1), 1-0 S(0), and 2-1 S(1), integrated over $\pm$1.2'' along the jet and centred on the star. The lines exhibit a peak at $-15$$\pm$4 km s$^{-1}$ from the systemic velocity. We believe this is not a spurious wavelength shift due to uneven illumination of the slit (cf. Bacciotti et al. 2002), since (1) the selected slit width of 0.3'' is much smaller than the seeing (0.7''--0.8''), and (2) the spectra at different slit positions show the same velocity within the uncertainty of the wavelength calibration.
The FWHM of 35 km s$^{-1}$ in the 1-0 S(1) profile is comparable to the spectral resolution ($\sim$30 km s$^{-1}$), implying that the internal velocity dispersion is much smaller. 
Assuming a $K$ magnitude of 6.74 (Muzerolle et al. 1998), we derive the 1-0 S(1) and 1-0 S(0) fluxes of 1.8$\pm$0.1$\times10^{-14}$ and 3.7$\pm$0.5$\times10^{-15}$ erg s$^{-1}$ cm$^{-2}$, respectively, and a 3-$\sigma$ upper limit of the 2-1 S(1) flux of 2.2$\times10^{-15}$ erg s$^{-1}$ cm$^{-2}$. The estimated 1-0 S(1) flux is consistent with the upper limit for the 1-0 S(1) flux ($<$$5.3\times10^{-14}$ erg s$^{-1}$ cm$^{-2}$) obtained by Carr (1990). 

\begin{figure}
  \begin{center}
    \psfig{file=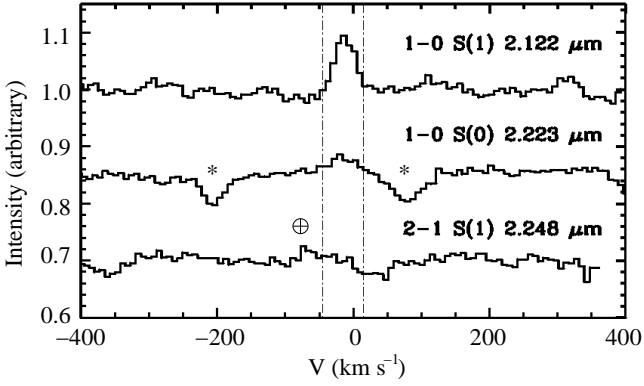,clip=,width=8.5cm}
  \end{center}
  \caption{Spectra at 1-0 S(1), 1-0 S(0), and 2-1 S(1) obtained along the jet. The P-V diagram obtained along the jet is integrated over $\pm$1.2'' of the continuum position, and normalized to the continuum at 2.12 $\mu$m.
The dot-dashed lines indicate the velocity range in which we measure the line flux. The asterisks indicate photospheric absorption features. Relative line fluxes between the spectra are conserved since the spectrograph covers all the lines with a single exposure.}
  \label{profiles}
\end{figure}

Table 1 shows the 1-0 S(0)/1-0 S(1) and 2-1 S(1)/1-0 S(1) ratios measured in the same region as for Fig. 3, i.e., within $\pm$1.2'' along the jet axis centred on the star.
For comparison, we also tabulate the line ratios for local thermal equilibrium (LTE) at $T=$1000--3000 K, where we have calculated these using the $A$-coefficients provided by Turner, Kirby-Docken, \& Dalgarno (1977). The table shows that observed line ratios are consistent with thermal excitation with a temperature of $\sim$2000 K. Such a spectrum is often observed in shocks in extended outflows and Herbig-Haro objects (see e.g., Gredel 1994; Eisl\"offel, Smith, \& Davis 2000b), although this could also result from thermal collisions in a dense UV-photodissociation region (see e.g., Sternberg \& Dalgarno 1989; Burton, Hollenbach, \& Tielens 1990) or via X-ray heating (see e.g., Maloney, Hollenbach, \& Tielens 1996; Tin\'e et al. 1997). We investigate this in detail in Sect. 3.3.

\begin{table}
 \centering
 \caption{H$_2$ Line Ratios$^a$}
 \begin{tabular}{@{}ccccc@{}}\hline \hline
                           & observed & \multicolumn{3}{c}{calculated}\\
                           &          & ($T$=1000 &2000 &3000 K)\\ \hline
$I_{2-1 S(1)}/I_{1-0 S(1)}$& $<$0.11$^b$ & 0.005 & 0.082 & 0.210 \\
$I_{1-0 S(0)}/I_{1-0 S(1)}$& 0.20$\pm$0.03  & 0.269 & 0.211 & 0.194 \\ \hline
\end{tabular}\\
$^a$ Extinction is not corrected.\\
$^b$ An upper limit for the 3-$\sigma$ level.
\end{table}

The blueshifted line profile and spatial offset away from the star indicate that the detected H$_2$ emission originates from the outflowing gas. The absence of red-shifted components is attributed to the obscuration of the counterflow by a circumstellar disk, common for forbidden line emission (e.g., Appenzeller, Jankovics, \& \"Ostreicher 1984; Edwards et al. 1987). The radial velocity of $-$15 km s$^{-1}$ is slower than the forbidden line outflow, which exhibits two blueshifted peaks at $\sim$200 and $\sim$20--100 km s$^{-1}$ (e.g., Edwards et al. 1987; Hamann 1994; Hartigan et al. 1995; Takami et al. 2002; Pyo et al. 2003). This trend is also observed in some Class I protostars, as shown by Pyo et al. (2002) and Davis et al. (2003). The measured flow width of $\sim$0.6'' in H$_2$ emission is wider than the forbidden line outflow, which has FWHM widths of $<$0.1'' and $\sim$0.2'' in high and low velocity components, respectively, at the same distance from the star (Bacciotti et al. 2002). These results indicate that H$_2$ emission originates from a different (i.e., slower and wider) flow component to the forbidden line outflow. 

Table 2 summarises the flow parameters of the warm H$_2$ wind.
The wind is not fully resolved in our spectra, thus the flow scale along the jet axis is estimated using the measured spatial offset of 0.2'', and assuming an inclination angle the same as the jet (45$^\circ$ --- Pyo et al. 2003).
The derived spatial scale of $\sim$40 AU is about half as large as that perpendicular to the jet axis ($\sim$80 AU), indicating that the flow has an opening angle of $\sim$90$^\circ$ if we assume zero width at the base. The luminosity is obtained assuming the extinction of $A_K$=0.2, based on the visual extinction to the object of $A_V$=2.2 (Muzerolle et al. 1998) and $A_K$$\sim$0.1$A_V$ (Becklin et al. 1978). The total H$_2$ mass is estimated assuming $M_{1-0 S(1)} / M_{H_2} =  1.28 \times 10^{-2}$ at $T=2000$ K, based on Smith \& Mac Low (1997) and Davis et al. (2001)\footnote{Assuming a population fraction $\chi_{v,j}(T)$=$g_{v,J} Z^{-1}(T)\times$\\exp$(-\Delta E_{v,J}/kT)$, where $g_{v,J}$ is the statistical weight, $\Delta E_{v,J}$ is the energy difference from the ground state, $Z(T)$ is the partition function $Z(T)$=0.024$T/(1-$exp$(-6000/T))$, and $T$ is the temperature. 
}. This mass should be a lower limit since the flow may include colder gas, which does not contribute to ro-vibrational H$_2$ emission. Indeed, the presence of such cold gas is predicted by shock (see Eisl\"offel et al. 2000b, and references therein) and UV-photodissociation models (e.g., Burton et al. 1990). The mean H$_2$ number density is estimated assuming a conical geometry with a filling factor of 1. This value should also be a lower limit due to the underestimated H$_2$ mass, and the fact that the filling factor may be less than 1 due to a clumpy or hollow geometry.

\begin{table}
 \centering
 \caption{H$_2$ Outflow Parameters}
 \begin{tabular}{@{}ll@{}} \hline \hline
Mean velocity$^a$ & $\sim$20 km s$^{-1}$\\
Flow scale along the jet axis$^a$ & $\sim$40 AU\\
Flow scale perpendicular to the jet axis & $\sim$80 AU\\
1-0 S(1) luminosity & $1.3 \times 10^{-5} ~L_{\odot}$ \\
H$_2$ mass emitting 1-0 S(1) & $2.7 \times 10^{-10} ~M_{\odot}$\\
Total H$_2$ mass$^b$ & $\ga 2.1 \times 10^{-8} ~M_{\odot}$\\
Mean H$_2$ number density$^b$     & $\ga4$$\times$$10^4$ cm$^{-3}$\\
Mass loss rate$^b$               & $\ga 2.2 \times 10^{-9} ~M_{\odot}$ yr$^{-1}$\\
Momentum flux$^b$  & $\ga 4.4 \times 10^{-8} ~M_{\odot}$ km s$^{-1}$ yr$^{-1}$\\
Kinematic luminosity$^b$ ($M_{H_2} V^3/2R$) & $\ga 7.2 \times 10^{-5} ~L_{\odot}$ \\ \hline
\end{tabular}\\
$^a$ Corrected for inclination to the line of sight.\\
$^b$ The lower limit is derived assuming the local thermal equilibrium at $T=$2000 K. See text for details.
\end{table}

\section{Discussion}
\subsection{Comparison with H$_2$ emission in other YSOs}
Unlike CO and CS, the H$_2$ molecule has no net dipole moment, hence it is not readily observed in ambient cold ($\ll$100 K) molecular clouds. However, shocks or UV/X-ray radiation in YSO allows the molecule to be excited to ro-vibrational states, thereby allowing us to probe active circumstellar regions.

In the last decades, a number of molecular outflows and Herbig-Haro objects have been found to exhibit near-infrared H$_2$ emission (see., e.g., Gautier et al. 1976; Gredel 1994; Davis et al. 1998; Chrysostomou et al 2000; Eisl\"offel et a. 2000b). Their morphology, line profiles, and line flux ratios are well explained by thermal excitation from shocks. In addition to the extended jets/outflows, their driving sources (Class I protostars) also exhibit near-infrared H$_2$ emission (Carr 1990; Greene \& Lada 1996; Reipurth \& Aspin 1997). Recent high-resolution spectroscopy and Fabry-Perot observations have confirmed that the emission originates from the base of the jet (Molecular Hydrogen Emission-Line Regions --- Davis et al. 2001, 2002). Their line profiles show properties similar to forbidden line emission in T Tauri stars: (1) the profiles often contain low (5--20 km s$^{-1}$) and high velocity (50--150 km s$^{-1}$) components, (2) the low-velocity component (LVC) is more common than the high-velocity component (HVC), and (3) the HVC is further offset than the LVC.

In addition to Class I protostars, several T Tauri stars are known to exhibit ro-vibrational H$_2$ emission. T Tau exhibits bright H$_2$ emission up to $\sim$1000 AU scale, and spectro-imaging by Herbst et al. (1996, 1997) show that the emission originates through interaction between outflowing and ambient gas. In contrast, H$_2$ emission in other T Tauri stars is likely to be associated with quiescent gas in the circumstellar disk. Weintraub, Kastner, \& Bary (2000) and Bary, Weintraub, \& Kastner (2002, 2003) show that the 1-0 S(1) line profile in these objects are centered at the star's systemic velocity, in contrast to the blueshifted emission of Class I protostars. The observed line widths of 9--14 km s$^{-1}$ and a double-peak profile in one of the objects (LkCa 15) are well explained by Keplerian motion in the circumstellar disk. The authors suggest that the H$_2$ molecules in these objects are excited by X-ray or UV radiation.

As shown in Sect. 2, near-infrared H$_2$ emission in DG Tau is associated with outflowing gas within 100 AU of the star, sharing the same origin as Class I protostars rather than T Tauri stars. This result is attributed to its evolutionary phase. Although DG Tau exhibits optical properties similar to T Tauri stars, its flat spectral energy distribution (Adams, Emerson, \& Fuller 1990) and energetic jet (e.g., Eisl\"offel \& Mundt 1998; Dougados et al. 2000; Davis et al. 2003) indicate that the object is in a transient phase between a Class I protostar and a T Tauri star (see, e.g., Greene \& Lada 1996 for discussion). The 1-0 S(1) luminosity in the DG Tau wind is 10--100 times fainter than Class I protostars (typically 10$^{-4}$--10$^{-3}$ $L_\odot$ --- Davis et al. 2002), and unlike some Class I protostars, a high-velocity component is absent. This could simply be a consequence of the evolution of DG Tau from the Class I to the Class II phase.

Bary et al. (2003) detected quiescent 1-0 S(1) emission in GG Tau A, TW Hya, LkCa 15, and DoAr 21, and the measured fluxes in these objects correspond to luminosities of 4.2$\times$10$^{-6}$, 1.1$\times$10$^{-7}$, 1.0$\times$10$^{-6}$, and 9.1$\times$10$^{-6}$ $L_\odot$, adopting distances of 140, 60, 140, and 140 pc, respectively (Wichmann et al. 1998; de Geus \& Burton 1991). These luminosities are smaller than the measured 1-0 S(1) luminosity in the DG Tau wind, typically at least by a factor of 3. Thus, 1-0 S(1) emission from the quiescent gas could contribute to the total 1-0 S(1) luminosity in DG Tau, but the total 1-0 S(1) luminosity should still be dominated by emission from the wind, consistent with the observed line profiles and offset shown in Sect. 2.


\subsection{Driving mechanism and wind launching region}
Observations of a wide-angled wind will allow us to investigate the driving mechanism of the jet/wind in detail. In particular, the low extinction of DG Tau allows us to compare the kinematics of the wind with the optical forbidden outflow, thereby allowing us to investigate the entire flow structure close to the driving source.

The {\it Hubble Space Telescope} observations by Bacciotti et al. (2000, 2002) revealed the detailed kinematics in the DG Tau jet with a resolution of $\sim$0.1''. The authors show that the centroidal velocity of the low velocity component monotonically decreases with distance from the jet axis, exhibiting an ``onion-like'' kinematic structure. At 0.2'' from the star, the component shows a centroidal velocity of $\sim$60$\pm$20 km s$^{-1}$ at the axis, and $\sim$30$\pm$10 km s$^{-1}$ 0.2'' from the axis (see Fig. 1 of Bacciotti et al. 2002).
 If we extrapolate the velocity field in the forbidden line jet to the typical width of the warm H$_2$ wind (i.e., $\pm$0.3'' from the axis), we could expect a velocity of $\sim$15 km s$^{-1}$, in excellent agreement with the measured velocity in H$_2$ emission.
This indicates that the warm molecular wind in H$_2$ emission can be attributed to an outer extension of the onion-like kinematic structure observed in the forbidden line outflow.
 This agrees with the magneto-centrifugal wind models, which predict that (1) fast collimated components and slow poorly-collimated components coexists (e.g., Shu et al. 2000), and (2) forbidden line emission traces only a limited fraction of the entire flow close to the jet axis (Li \& Shu 1996; Cabrit et al. 1999; Shang et al. 2002).

In addition to the magneto-centrifugal wind scenario, the presence of the warm molecular wind may also be explained by entrainment of the surrounding material by the collimated jet. This mechanism has been proposed, for instance, for H$_2$ emission in Class I protostars (Davis et al. 2003) and the low velocity component of forbidden line emission (Pyo et al. 2003). Although our data do not exclude this possibility, it may have a few problems. Model calculations by Li \& Shu (1996) suggest that such interaction close to the driving source is weak because of ``magnetic cushioning'', i.e., the effects of frozen-in magnetic field in the flow. In addition, $^{13}$CO observations by Kitamura, Kawabe, \& Saito (1996) suggest that the majority of the matter surrounding the jet has already been blown away, leaving nothing to entrain. These authors also show that the remnant, flat envelope at a $\sim$1000 AU scale is slowly ($\sim$1.5 km s$^{-1}$) expanding, and it may be difficult to understand such motion without invoking with a wide-angled wind.

Since the majority of the surrounding matter has already been blown away (Kitamura et al. 1996), the molecular wind must emerge from the circumstellar disk, i.e., the remaining reservoir of the H$_2$ molecules. This fact may give us a clue to the location of the jet/wind launching region, the position of which have been debated for over a decade. Magneto-centrifugal wind models suggest two possibilities: (1) the X-wind model predicts that the flow emerges between the inner disk edge and stellar magnetosphere within a few stellar radii (Shu et al. 2000); and (2) the disk wind model predicts that the flow emerges from a wider ($>$1 AU) disk surface (e.g., K\"onigl \& Pudritz 2000; Ferreira 1997). The disk wind model can easily explain the presence of H$_2$ molecules in the wind, whereas it is not clear whether the X-wind model can do so. This is because any H$_2$ molecules are likely to become dissociated close ($<$0.1 AU) to the star because of the harsh environment. Further modelling including molecular chemistry is necessary to investigate this in detail.

\subsection{Excitation and heating}
Understanding the heating in the jet/wind is desirable, in particular in close proximity to the driving source, since it will allow us to perform comparisons with kinematic models and observations, and investigate their driving mechanism in detail. In the case of the collimated jet, several heating mechanisms have been proposed, including shocks (e.g., Hartigan et al. 1995), turbulent dissipation in a viscous mixing layer (e.g., Binette et al. 1999), ambipolar diffusion (Safier 1993; Garcia et al. 2001), and X-rays from the star (Shang et al. 2002). Recent high-resolution observations (e.g., Bacciotti et al. 2000; Dougados et al. 2000) and detailed analysis of optical forbidden lines (Lavalley-Fouquet, Cabrit, \& Dougados et al. 2000; Dougados, Cabrit, \& Lavalley-Fouquet 2002) indicate that shocks in the form of internal working surfaces are likely to be the cause of heating on 10--100 AU scales. This conclusion is corroborated by the detection of HeI emission in the DG Tau jet (Takami et al. 2002). On the other hand, Takami et al. (2001, 2003) revealed the presence of H$\alpha$ jets on a few AU scales from some T Tauri stars. Based on their spatial scale and line profile, the presence of another heating mechanism in the inner ($<$10 AU) region is suggested.

What heats and excites molecular hydrogen in the warm wind? 
The cooling timescale of molecular gas  at $\sim$2000 K is less than a year (see Hollenbach \& Natta 1995), smaller than the dynamical scale of the warm molecular wind we observed ($\sim$10 yr). This indicates the presence of a heating mechanism, which keeps the wind warm and allows H$_2$ emission to be observed. 
Possible mechanisms include: (1) shocks, which is responsible for near-infrared H$_2$ emission in the extended jet/outflows
(e.g., Gredel 1994; Davis et al. 1998; Chrysostomou et al 2000; Eisl\"offel et al. 2000b);
(2) UV fluorescence, which is responsible for UV H$_2$ emission in accreting T Tauri stars (Valenti, Johns-Krull, \& Linsky 2000; Ardila et al. 2002); (3) X-ray, which may be responsible for near-infrared H$_2$ emission in the circumstellar disk of T Tauri stars (Weintraub et al. 2001; Bary et al. 2002, 2003); and (4) ambipolar diffusion (Safier 1993).

If the warm molecular wind consisted of a dense UV-photodissociation region (PDR), we could expect a 1-0 S(1) surface brightness of up to $\sim$10$^{-3}$ erg s$^{-1}$ cm$^{-2}$ str$^{-1}$ based on (1) the UV continuum flux from the star ($\sim$2$\times$10$^{-13}$ erg s$^{-1}$ cm$^{-2}$ \AA$^{-1}$ at 1400--2000 \AA~--- Gullbring et al. 2000), (2) the spatial scale of the warm molecular wind, and (3) PDR models by Burton et al. (1990). This would require a surface area of at least $\sim$4$\times$10$^{30}$ cm$^2$ to account for the observed 1-0 S(1) luminosity. Together with a typical column density of the H$_2$ emitting region ($\sim$10$^{21}$ cm $^{-2}$ --- Burton et al. 1990) and the flow velocity shown in Table 2, we derive a lower limit of the momentum flux of 7$\times$10$^{-6}$ $M_\odot$ km s$^{-1}$ yr$^{-1}$. This flux is extremely large for the evolutionary phase of this object: it is even much larger than a typical momentum flux of molecular outflows associated with its younger progenitors, i.e.,  Class I protostars (2$\times$10$^{-6}$ $M_\odot$ km s$^{-1}$ yr$^{-1}$ --- Bontemps et al. 1996).
We then conclude that UV excitation is not likely for the heating of the warm molecular wind.

X-ray photodissociation region (XDRs) models by Maloney et al. (1996) show that the X-ray surface brightness is a function of the X-ray luminosity at the radiation source, distance to the XDRs, column density, and hydrogen nucleus density. Based on their models, we derive a 1-0 S(1) surface brightness of 10$^{-6}$ erg s$^{-1}$ cm$^{-2}$ str$^{-1}$ providing (1) a typical X-ray luminosity of a classical T Tauri star of 10$^{29}$ erg s$^{-1}$ (Neuh\"auser et al. 1995); (2) hydrogen nucleus density of 10$^5$ cm$^{-3}$; (3) a column density of 10$^{21}$ cm$^{-2}$, the same density as assumed for UV excitation described above. The derived surface brightness described above is smaller than that for UV excitation by a factor of $\sim$10$^3$, implying that a larger surface area or column density would be required to account for the observed 1-0 S(1) luminosity. However, such assumptions would provide even larger flow momentum flux than we estimated for UV excitation.  We conclude that neither UV or X-rays constitute the primary mechanisms.
Note that the hydrogen nucleus density in the warm molecular wind may be higher than we assumed above (see Table 2), although a higher density would not increase the 1-0 S(1) luminosity. This is because it would cause a larger cooling rate, thereby allowing X-ray heating to be less effective. 

In contrast, shocks are a likely mechanism for the heating of the warm molecular wind, since they are observed in the jet/outflows of various YSOs. The flow velocity of $\sim$20 km s$^{-1}$ in the DG Tau wind is much larger than the sound velocity (3 km s$^{-1}$ at $T$=2000 K), indicating that shocks could occur in the same manner as for extended jets and outflows: i.e., as internal working surfaces (e.g., Chrysostomou et al. 2000; Davis et al. 2000) or supersonic turbulence (Davis \& Smith 1996). In addition, the observed line flux ratios shown in Table 1 are similar to those in shock-excited extended outflows, which indicate thermal temperatures of 1000--4000 K (see., e.g., Gredel 1994; Eisl\"offel et a. 2000b), supporting the shock heating scenario.

In addition to shocks, ambipolar diffusion may also explain the heating in the warm molecular wind we observe. Safier (1993) shows that H$_2$ molecules could survive in the outer region ($\ge$1 AU from the flow axis) of an onion-like kinematic structure, and they could be heated up to a few thousands of Kelvins, consistent with our results. He also predicts that the temperature in the wind increases with the distance. The inner structure of the observed wind could be resolved by use of adaptive optics on 8--10 m telescopes, and such observations in the near future would allow us to identify the heating mechanisms, testing the validity of existing models.

%
%
%
%
%
%
%
%
%
%

%
%
%
%



\section{Conclusions}
We detected near-infrared H$_2$ emission in DG Tau, one of the most active T Tauri stars known. The spatial extension along and perpendicular to the jet, and measured blueshift of 15 km s$^{-1}$, indicate that the emission originates from a warm molecular wind with a flow length and width of $\sim$40 and $\sim$80 AU, respectively. Via detailed comparison of the spatial scale and velocity field, we conclude that H$_2$ and forbidden line emission originate from different regions of the entire flow, which includes (1) fast, well-collimated and partially ionised streamlines, and (2) slow, poorly-collimated, and molecular streamlines. Such a flow geometry agrees with predictions of magneto-centrifugal disk wind models. Although it could also be explained by the entrainment of the surrounding material by the collimated jet, we consider it unlikely.

The measured line flux ratios ($I_{1-0 S(0)}/I_{1-0 S(1)}$ and the upper limit of $I_{2-1 S(1)}/I_{1-0 S(1)}$) indicate that the flow is thermalized at a temperature of $\sim$2000 K. The flow is likely to be heated by shocks or ambipolar diffusion.

\begin{acknowledgements}
We are grateful to the referee (Dr. J. Eisl\"offel) for his providing useful comments.
We acknowledge
the data analysis facilities provided by the Starlink Project which
is run by CCLRC on behalf of PPARC.
MT thanks PPARC for support through a PDRA.
This research has made use of the Simbad Database, operated at CDS. Strasbourg, France, and of the NASA's Astrophysics Data System Abstract Service.
\end{acknowledgements}

\end{document}